\documentstyle[11pt,AATS]{article}		

\begin{document}	

\title{Time at the Beginning}

\author{Michael S. Turner}
\affil{Departments of Astronomy \& Astrophysics and of Physics\\
Enrico Fermi Institute, The University of Chicago\\
Chicago, IL~~60637-1433\\}
\bigskip
\affil{NASA/Fermilab Astrophysics Center\\
Fermi National Accelerator Laboratory\\
Batavia, IL~~60510-0500}

\begin{abstract}
Age consistency for the Universe today has been
an important cosmological test.  Even more powerful consistency tests at
times as early as $10^{-32}\,$sec lie ahead in the precision
era of cosmology.  I outline tests based upon cosmic microwave
background (CMB) anisotropy, big-bang nucleosynthesis (BBN),
particle dark matter, phase transitions, and inflation.  The ultimate
cosmic timescale -- the fate of the Universe -- will be
in doubt until the mystery of the dark energy is unraveled.
\end{abstract}


\section{Introduction}

The cosmic clock ticks logarithmically.  The Universe's past
can be divided into three well defined epochs:
the quantum era ($10^{-42}\sec$ to $10^{-22}\sec$);
the quark era ($10^{-22}\sec$ to $10^{-2}\sec$); and
the hot big-bang era ($10^{-2}\sec$ to $10^{17}\sec$).
Earlier than the quantum era is the Planck era ($t<10^{-42}\,$sec);
it holds the key to understanding the birth of the Universe,
but requires a quantum theory of gravity to proceed.
Just recently the Universe has entered a new era, of undetermined
duration, where dark energy
and its repulsive gravity are dominating the dynamics
of the Universe (more later).

The third era derives its name from the very successful
cosmological model that describes it; the important
events include the synthesis of the light elements,
the last scattering of radiation, the formation of
large-scale structure and the onset of accelerated
expansion.  The earlier two eras are still largely
terra incognita, though we have some tantalizing ideas:
the transition from quark/gluon plasma to hadronic
matter, the electroweak phase transition, and the birth
of particle dark matter during the quark era; and
inflation and the origin of the matter -- antimatter
asymmetry during the quantum era (see Figure).

There is no doubt that cosmology is in the midst of the
most exciting period ever.  Progress in understanding
the origin and evolution of the Universe is proceeding
at a stunning pace.
I mention a few of the highlights
of the past decade:  COBE discovery of the small density
inhomogeneities ($\delta \rho /\rho \sim 10^{-5}$) that
seeded large-scale structure, the mapping of CMB anisotropy on
scales down to 0.1 degree and the determination of the
shape of the Universe (flat!), identification of the
epoch of galaxy formation, determination of the
Hubble parameter to a precision of 10\%,
and the discovery of cosmic accelerated expansion.

There is much more to come.  These recent discoveries
and those still to be made will test the framework
of the standard cosmology and begin to open the
quark and quantum eras.  We are already on the way
to establishing a new standard cosmology:
a flat Universe comprised of 2/3rds dark energy and
1/3rd dark matter that is accelerating today and
whose seeds for structure came from quantum
fluctuations stretched to astrophysical size during
inflation.  Much remains to be done to put the
new cosmology on the same firm footing as the
hot big-bang model; much of cosmology over the
next two decades will be devoted to this (see e.g., Turner, 2001a).

The richness and redundancy of the measurements to be made
will also allow a number of
consistency tests of the big-bang framework and
its theoretical foundation, general relativity.
Many of these tests involve cosmic timescales, and
that is the subject of my talk.

\section{$H_0t_0$:  ``Sandage Consistency Test''}

The product of the present age and Hubble constant is
a powerful consistency test.
The standard hot big bang has not always
cleared this hurdle, cf. the late 1940s when the Hubble age
of about 2 billion years fell short of the
age of the solar system by a factor of two (see e.g.,
Kragh, 1996) or the more recent scares when
some measurements of $H_0$ and $t_0$ drifted upward.

Recognizing the importance of age consistency,
Sandage has devoted much of his career to measuring the
Hubble constant $H_0$ and independent indicators of the
age of the Universe (and pioneered
many of the methods); hence the title of this section.

The nature of the test has changed over Sandage's
career; the importance has not.  Until recently
one would have written $H_0t_0$ in terms of one
parameter, $\Omega_M$, the fraction
of critical density in matter, also assumed to be
the total fraction of critical density contributed
by all forms of matter and energy ($\equiv \Omega_0$).  
The discovery of
accelerated expansion increased the number
of parameters to two, $\Omega_M$ and $\Omega_\Lambda$
(fraction of critical density in a cosmological
constant; $\Omega_0 = \Omega_M + \Omega_\Lambda$).
The realization that accelerated expansion is here
to stay, while the cosmological constant may not,
added another parameter, $w = p_X/\rho_X$ (the equation-of-state
of the dark energy) (see below).  The determination
from CMB anisotropy measurements that the Universe
is flat, effectively reduced the number of parameters to two,
$\Omega_M$ and $w$:
\begin{eqnarray*}
H_0t_0 & = & f(\Omega_0)\qquad\qquad\qquad\ \ \ \, {\rm \rightarrow 1998} \\
        & = & f(\Omega_M,\Omega_\Lambda ) \qquad\ \  {\rm 1998 \rightarrow 1999} \\
        & = & f(\Omega_M, \Omega_X, w) \ \ \ \ {\rm 1999 \rightarrow 2000} \\
        & = & f(\Omega_M, w) \qquad\ \ \ \, {\rm 2000 \rightarrow\ \ \ ?} \\
        & = & f(w)   \qquad\qquad\ \ \ \ \ \, {\rm ?\ \ \  \rightarrow\ \ \ ?}
\end{eqnarray*}
For the first two cases, there are well known analytic formulae
for $f$; for the others there are not.

There is hope, that in the next ten years $\Omega_M$
will be pinned down by a combination of CMB, cluster
and large-scale structure measurements, reducing the
number of parameters to one again.  At the moment,
my analysis of this data gives, $\Omega_M = 0.33\pm 0.035$
(Turner, 2001b), and I believe there will be significant
improvement over the next decade.

The function $f$ that accounts for the effect of slowing/speeding up
on the age of the Universe is given by (for a coasting
universe $f= 1$):
\begin{eqnarray}
f \equiv H_0t_0 & = & \int_0^\infty {dz \over (1+z)H(z)/H_0} \nonumber \\
H(z)/H_0 & = &  \left[\Omega_M(1+z)^3 + \Omega_X (1+z)^{3(1+w)} +
        (1-\Omega_0)(1+z)^2\right]^{1/2}
\end{eqnarray}
where the small contribution from the CMB and relativistic
neutrinos ($\Omega_R\sim 10^{-4}$) has been neglected and
$w$ has been assumed to be constant.

To illustrate the status of this test,
for $H_0$ and $t_0$ I will use the values
reported at this meeting, $H_0 = 72 \pm 7 \,
{\rm km\,sec^{-1}\,Mpc^{-1}}$ (Freedman, 2001) and $t_0 =
13.5 \pm 1.5 \,$Gyr (Chaboyer, 2001).  This leads to
$$ H_0t_0 = 0.94 \pm 0.14$$
Taking $\Omega_0 = 1$ and $\Omega_M \simeq 0.35$, this is consistent
with $w < -{1\over 3}$ (at $1\sigma$) -- the new cosmology
passes the Sandage consistency test with flying colors!
While not a tight constraint on $w$,
it does provides more evidence for dark energy (i.e., a smooth
component with large, negative pressure):  if the dark energy
were pressureless ($w=0$), then $H_0t_0 =2/3$, which is
clearly inconsistent with current data.

At present, the Sandage test has a precision of about 15\%; what
kind of improvement could one hope for over the next decade.
Together, physics-based measures of $H_0$ (especially the CMB)
and distance scale measures might well pin down the Hubble constant
to a few percent.  However, it is difficult to imagine independent
measures of the age achieving similar accuracy.  One irreducible
uncertainty for all methods, is the time to formation (of stars,
globular clusters, white dwarfs, etc).  Reducing this uncertainty
to 0.5\,Gyr would still leave a 5\% uncertainty in $H_0t_0$.  As
I will discuss, other tests of age consistency, at much earlier
times, are likely to be significantly more precise.

\subsection{Mapping the expansion rate}

The expansion rate at any epoch is related to the age of the
Universe.  For example:  during a matter-dominated epoch, the
scale factor $R(t) \propto t^{2/3}$ and $t = 2/3H(t)$;
and during a radiation-dominated epoch $R(t) \propto t^{1/2}$
and $t = 1/2H(t)$.  The expansion rate is more accessible than
the age at early times.

The expansion rate determines a key observable:  the comoving
distance to an object a redshift $z$,
\begin{equation}
H_0r(z) = \int^z_0 {dz \over H(z)/H_0}
\end{equation}
where a flat Universe has been assumed.  The cosmological volume
element, luminosity distance and angular-diameter distance are
all directly related to $r(z)$
\begin{eqnarray*}
dV/dzd\Omega & = & r(z)^2/H(z) \\
d_L & = & (1+z)r(z) \\
d_A & = & r(z)/(1+z)
\end{eqnarray*}

A number of probes of the expansion history from $z=0$ to
$z\sim 2$ have been discussed recently.  They include
gathering a large sample of type Ia supernovae (SNAP) and
counting halos or clusters of galaxies.  From these
``experiments,'' individually or taken together, one
could imagine mapping out $H(z)$ back to redshift $z\sim 2$
(see e.g., Huterer \& Turner, 2000; or Tegmark, 2001).
Note that the volume element and luminosity distance
taken together, can in principle directly yield $H(z)$.

Loeb (1998) has gone one step further, suggesting ultra-precise
redshift measurements of the Lyman-alpha forest over a decade or more
could be used to determine $H(z)$ directly(!).  The idea
is to measure the tiny time variation of the redshifts of
thousands of Lyman-alpha clouds,
\begin{eqnarray}
\delta z & = & [ -H(z_1) + (1+z_1)H_0 ] \delta t \nonumber \\
         & \sim & 0.2\,(\delta t /10\,{\rm yrs})\,{\rm m\,sec^{-1}}\qquad
         {\rm for\ }z=3
\end{eqnarray}
where $\delta t$ is the time interval of the observations.
Whether or not this bold idea can be carried out remains to
be seen, but it certainly is exciting to think about.

\section{CMB anisotropy and acoustic peaks}

As is now very familiar, the power spectrum (in multipole
space) of CMB anisotropy arises due to acoustic oscillations
in the baryon -- photon fluid around the time of last
scattering (see e.g., Hu et al, 1997).  At this time, the
baryons (and electrons) are falling into the dark-matter
potential wells; still coupled to photons (through Thomson
scattering off electrons), the infall is resisted by photon
pressure and oscillations ensue.  At maximum compression
the photons are heated and at maximum rarefaction they
are cooled.

The CMB is a snapshot of the Universe at 400,000 yrs; regions
caught at maximum compression (rarefaction) lead to hot
(cold) spots on the microwave sky.  While the
physics is most easily explained in real space, the signal
is best seen in multipole space, as a series of peaks
and valleys.  The power at multipole $l$ is
largely due to $k$-modes satisfying:  $k \sim lH_0$

The condition for maxima in the power spectrum is:
$\omega_n t_{\rm LS}/\sqrt{3} = n\pi$, where $n =1,2,3
\cdots$; the odd peaks are compression peaks
and the even peaks are rarefaction peaks, $\omega
= (1+z_{\rm LS})k$ is the physical oscillation
frequency at the time of last scattering,
and $t_{\rm LS} \simeq 400,000\,$yrs is
the age of the Universe then.  Thus,
in a flat Universe, the harmonic peaks occur at multipoles
$$ l_n \sim {n\pi \over H_0t_{\rm LS} (1+z_{\rm LS})} \sim 200 n$$

While the above formulae are approximate, they
capture the physics.  In particular, that the positions of
the peaks depends upon the age of the Universe at
last scattering.  There will be sufficient redundancy in
the information encoded in the 3000 or so multipole
amplitudes that will be measured to not only determine
cosmological parameters, but also to check the consistency
of the standard relationship between age and energy density.

In particular, an analysis by Lopez et al
(1999) has projected that the Planck mission will be
able to peg the neutrino contribution to the energy
density of the Universe at last scattering to about
1\%.  The expansion rate is related to the energy density,
$H^2 =8\pi G\rho /3$, and neutrinos contribute about
1/5th of the total energy density at the time of last
scattering.  A quick estimate from Lopez et al (1999)
indicates that ultimately, CMB anisotropy will provide
an age consistency test at about 0.1\% precision,
400,000 yrs after the beginning.

The limitations of this test should be noted however.
Actually determining the age of the Universe at last
scattering, whose redshift is readily determined by
thermodynamic considerations
($1+z_{\rm LS}) \simeq 1100$), is pegged to the
present Hubble constant, $t_{\rm LS} \simeq
H_0^{-1}/[\Omega_M(1+z_{\rm LS})^{3/2}$, and thus
can be no more accurate than the age of the Universe itself.
The CMB consistency above actually probes the relationship
between expansion rate and the energy density of
the Universe, a test of the big-bang framework and general relativity.

\section{Big-bang Nucleosynthesis}

Big-bang nucleosynthesis (BBN) is a cosmological experiment
of great importance.  In essence, it is a quenched
nuclear reactor whose by-products are the
primeval mix from which the first generation stars were born.

The expansion rate of the Universe controls the quench
rate:  $|\dot T/T| = H$.  The yields of D, $^3$He, $^4$He
and $^7$Li, the primary products of BBN, are sensitive to
the quench rate, and thus to the expansion rate
\begin{equation}
{\delta X\over X} \sim {\delta H \over H_{\rm STD}}
\end{equation}
where the coefficient of proportionality varies from
about $1.5$ for $^7$Li to about $0.6$ for $^4$He.

Measurements of the primeval deuterium abundance
(see e.g., O'Meara et al, 2001) together with
predictions for its BBN yield
pin down the baryon density, $\Omega_Bh^2 =
0.02\pm 0.001$ (Burles et al, 2001a).  Using this value,
the other light-element abundances can be predicted,
e.g., $Y_P = 0.2472\pm 0.0005$.
Of the four light elements, the primeval $^4$He abundance is known most
accurately, $Y_P = 0.244\pm 0.002$ (though there
is still much debate about systematic error; see e.g.,
Burles et al, 2001b, or Olive et al, 2000).

The sensitivity of the $^4$He abundance to the expansion
rate is:  $\delta Y_P = 0.15(\delta H/H_{\rm STD})$.  The
current agreement of prediction with observation translates
to a 2\% test of the consistency of the big-bang
prediction for the expansion rate around 1 sec, or about
5 times better than the Sandage test!  With improved
measurements of the baryon density, both from BBN and
CMB anisotropy (see below), one might hope to see an
improvement of a factor
of five or so in the next decade.

Finally, as Carroll \& Kaplinghat (2001) have recently emphasized,
BBN can be discussed without reference to the standard
cosmological model.  The yields can
be analyzed completely in terms of the quench rate, $|\dot T/T|$,
and nuclear input data with only the assumption of the Robertson --
Walker line element (and not the Friedmann equations which
relate the quench rate to the temperature).
BBN not only offers a window on the very early
Universe, but an almost model-independent timescale test.

\section{CMB + BBN and the Baryon Density}

Together, the CMB and BBN offer a remarkable test of the
consistency of the standard cosmology and general relativity.
As mentioned in the previous section, measurements of the
primeval deuterium abundance together with precise theoretical
predictions can be used to infer the baryon density:
\begin{equation}
\Omega_B({\rm BBN}) h^2 = 0.02 \pm 0.001
\end{equation}
Since the first measurements of the primeval deuterium abundance
in 1998, BBN has provided the best determination of the
baryon density (see e.g., Schramm \& Turner, 1998).

The CMB provides an independent measure of the baryon density
based upon the physics of gravity-driven acoustic oscillations
400,000 yrs after BBN.  Specifically, it is the ratio of the
heights of the odd to even acoustic peaks that is sensitive
to the baryon density (and insensitive to other cosmic
parameters; see e.g., Hu et al, 1997).  The
ratio of the heights of the first and second peaks is
\begin{equation}
{\rm peak_1 \over peak_2} \simeq 2 \left(\Omega_Bh^2 /0.02\right)^{2/3}
\end{equation}
where the peak heights are measured in microKelvin$^2$.

The Boomerang and Maxima experiments were the first CMB
experiments to probe the first and second peaks (and to
show that the Universe is flat; de Bernardis et al, 2000;
Hanany et al, 2000); based upon their results
a value for the baryon density was inferred (Jaffe et al, 2001)
$$\Omega_Bh^2 = 0.032^{+0.005}_{-0.004} $$
While this baryon density is about $2\sigma$ higher than
the BBN value, it confirmed a key prediction of BBN -- a
baryon density far below that of the best estimates of
the total matter density ($\Omega_M \sim 0.3$ vs. $\Omega_B \sim 0.05$).

Only a couple of months after this meeting, Carlstrom's DASI
experiment at the South Pole announced even more accurate
and independent measurements
of the first three acoustic peaks (Pryke et al, 2001)
and arrived at a slightly lower baryon density:
\begin{equation}
\Omega_B({\rm CMB}) h^2 = 0.022^{+0.004}_{-0.003}
\end{equation}
which is bang on the BBN value!  Boomerang analyzed more data
and refined their beam map and pointing model and arrived
at an identical value (Netterfield et al, 2001).

The agreement between these two numbers is stunning.  Using
the fact that (D/H)$_P \simeq 3\times 10^{-5} (\Omega_Bh^2/0.02)^{-1.6}$
and the sensitivity of D/H to the expansion rate,
$${\delta {\rm (D/H)} \over ({\rm D/H})}
\simeq 1.4 {\delta H\over H_{\rm STD} },$$
it follows that
$${\delta (\Omega_Bh^2) \over \Omega_Bh^2} \sim -0.9 {\delta H\over
H_{\rm STD}} $$
Thus, the agreement of the BBN and CMB baryon densities checks
the consistency of the standard expansion rate (at about 1\,sec)
to a precision of about 15\%.  Eventually, both determinations
of the baryon density should achieve about 1\%
or so accuracy, and a factor of ten in the precision of this test
might be expected.

\section{Particle Relics}

The evidence for particle dark matter has only gotten stronger:
firmer evidence for $\Omega_B \ll \Omega_M$ (discussed above); 
the many successes of the CDM scenario of structure formation
(and no viable model for structure formation without particle
dark matter); the detection of acoustic peaks as predicted
by inflation and CDM (Netterfield et al, 2001; Pryke et al,
2001); and growing circumstantial evidence for supersymmetry.

The most promising particle candidate is the lightest supersymmetric
particle, which in most models is a neutralino of mass 100 to
300 GeV (see e.g., Jungman et al, 1996).
Relic neutralinos remain numerous today because of
the incompleteness of
neutralino annihilations in the early Universe.

At temperatures when $kT \gg m_\chi c^2$, neutralinos and anti-neutralinos are
present in numbers comparable to that of photons; as the temperature
drops neutralinos must annihilate to maintain thermal abundance
(a factor $e^{-m_\chi c^2/kT}$ less than that of photons).  Eventually,
annihilations cannot keep pace with the quench rate (the annihilation
rate per neutralino, $\Gamma = n_\chi <\sigma v>_{\rm ann}$,
falls rapidly as the neutralino abundance decreases exponentially) and the neutralino
abundance ``freezes out.''   Freeze out occurs at a temperature,
$kT_f \sim m_\chi c^2 /30$, corresponding to a time of around
$10^{-7}\,$sec.  The mass density contributed by relic neutralinos
is given by
\begin{equation}
\Omega_\chi h^2 \ \ \propto\ \ {H(kT \sim m_\chi c^2/30 )
\over <\sigma v>_{\rm ANN}}
\end{equation}
While an approximation, this formula captures the essence of the
neutralino production process (see e.g., Kolb \& Turner, 1990).

Here is the future timescale test:  $\Omega_\chi h^2$ will be measured by
CMB anisotropy experiments to percent level precision.  The properties
of the neutralino can be measured at an accelerator lab (next linear
collider?) to 10\% precision (Brhlik et al, 2001).  From this, the expansion rate at
$kT = m_\chi c^2/30$ can be inferred and compared to the standard
formula.  (To be more precise, the measured properties of the neutralino
and the Boltzmann equation in the expanding Universe can be used to
predict the relic mass density and compared with the value inferred
from CMB measurements.)  If the neutralino is indeed the dark-matter particle
(or another particle that can be produced in the lab), we can look
forward to a 10\% consistency test of the expansion rate at a time
of less one microsecond!

\section{Phase Transitions}

If current ideas about particle physics are correct, then,
during its earliest moments, the
Universe should have gone through a series of phase transitions
associated with symmetry breaking (e.g., QCD, electroweak, grand
unification, compactification?).  During a first-order phase transition, the Universe
gets ``shaken up'' as bubbles of the new phase expand and collide.
This can lead to the production of prodigious amounts of gravitational
radiation, resulting in $\Omega_{\rm GW} h^2
\sim 10^{-10}$ today.  Laboratory-based knowledge of particle physics
(e.g., for the electroweak phase transition, the mass of the Higgs
boson) can be used to predict
with precision the spectrum and amount of gravitational radiation
(Kosowsky, Turner \& Watkins, 1992); both depend directly
upon the expansion rate at the temperature at which the phase transition
takes place.  If the stochastic background of gravitational waves
from a phase transition can be detected and the particle physics
independently probed in the laboratory a cosmological timescale test
can be carried out at very early times (e.g., for the electroweak
phase transition, $t\sim 10^{-11}\,$sec).

\section{Inflation}

Inflation also produces gravitational waves, by a different
mechanism -- deSitter-space produced quantum fluctuations
in the space-time metric.  The stochastic background of
gravity waves produced by inflation have wavelengths from
1\,km to the beyond the size of the present horizon.  Gravity
waves are one of the three key predictions of inflation (together
with a flat Universe and a nearly scale-invariant spectrum of
adiabatic, Gaussian density perturbations; see e.g., Turner, 1997a).

Detection of gravitational waves, either by their imprint on
CMB anisotropy and/or polarization or directly by a gravity-wave
detector (LIGO?, LISA?), would not only confirm a key prediction of
inflation, but would also reveal the timescale for inflation:
The level of gravitational radiation produced by inflation
is directly related to the expansion rate during inflation:
$$h_{\rm GW} \sim H_I/m_{\rm Pl} $$
Measuring the dimensionless metric strain $h_{\rm GW}$ gives
the Hubble parameter during inflation in units of the Planck
energy ($=1.22\times 10^{19}\,$GeV).  Further, if the spectrum
of gravitational radiation can be probed, then there is a consistency
test of inflation (and cosmology):  $T/S = -5n_T$, where $T$ ($S$)
is the contribution of gravity waves (density perturbations) to
the quadrupole CMB anisotropy, and $n_T$ is the spectral
index of the inflation-produced gravitational waves (Turner, 1997a,b).

As a practical matter, gravity waves from inflation can probably
only be detected if $H_I > 3\times 10^{12}\,$GeV, corresponding
to a timescale of $H_I^{-1} < 10^{-30}\,$sec.  Said another way, 
{\em if} gravity waves from inflation are detected
we will be probing the Universe at a time at least as early
as $10^{-30}\,$sec.

\section{Dark energy \& destiny}

For decades cosmologists have believed that geometry (or equivalently
$\Omega_0$) and the fate of the Universe were linked.  The
shape of the Universe has been determined through measurements
of CMB anisotropy,
$\Omega_0 = 1.0\pm 0.04$ (Pryke et al, 2001; Netterfield et al,
2001; Hanany et al, 2001), but we are further away than
ever from determining the destiny of the Universe.
This is because 2/3rds of the critical
density is in dark energy rather than matter, and
the connection between geometry and destiny only applies
when the Universe is comprised of matter (or more precisely,
stress-energy with $p > -\rho /3$; see Krauss \& Turner, 1999).

The discovery of accelerated expansion through type Ia
supernovae distance measurements (see Tonry, 2001)
was surprising,  but can easily be accommodated within the
framework of general relativity and the hot big-bang cosmology:
In general relativity, the source of gravity is $\rho + 3 p$,
so that a fluid that is very elastic (negative pressure comparable
in magnitude to energy density) has repulsive gravity.
The simplest example is the quantum vacuum (mathematically equivalent
to a cosmological constant), for which $p = -\rho$.

Unfortunately, all estimates of the energy of the quantum vacuum
exceed by at least 55 orders-of-magnitude what is required to
explain the acceleration of the Universe, suggesting to many
that ``nothing weighs nothing'' and that something else with
large negative pressure is causing the accelerated expansion (e.g.,
a rolling scalar field, aka as mini inflation or quintessence,
or a frustrated network of cosmic defects; see Turner, 2000 or
Carroll, 2001).  Borrowing from Zwicky, I have coined the term
``dark energy'' to describe this stuff, which is clearly nonluminous
and more ``energy like'' than ``matter like'' (since $|p|/\rho \sim 1$).
It seems to be very smoothly distributed and its primary effect is on
the expansion of the Universe.  The first handle we will have
in determining its nature is measuring its equation-of-state
$w=p/\rho$ through cosmological observations (see e.g., Huterer
\& Turner, 2000).

Until we figure out the nature of the dark energy, the ultimate
timescale question is on hold.  Accepting that the Universe is
flat (or at least that our bubble is; see Guth, 2001), the possibilities
for destiny are wide open:  continued accelerated expansion
(and the almost complete ``red  out'' of the extragalactic
sky in 150 billion years) if the dark energy is
vacuum energy; eternal slowing if the dark energy dissipates
and matter takes over the dynamics; or even recollapse if the
dark energy dissipates revealing a small, negative cosmological
constant (Krauss \& Turner, 1999).

\section{Concluding Remarks}

Cosmology is in the midst of a Golden Age (see e.g., Turner, 2001a).
As this meeting has illustrated, other areas of astronomy have not been 
left behind either.  Astronomy in general is in the midst of the most
exciting period of discovery ever.

The enormous variety and range of timescales in astrophysics
makes the subject rich.  Cosmology provides an especially good
illustration, because the cosmological clock ticks logarithmically.
Consistency checks on the cosmological clock have and will continue
to play an important role in validating the standard cosmological
model.  While the consistency of the present age and expansion
rate (Sandage test) is important and will improve in accuracy,
from its present 15\% to perhaps 5\%, there are many other
age consistency tests in cosmology, whose precision may well
approach a few tenths of a percent (e.g., BBN and CMB), and
extend to times as early as $10^{-30}\,$sec.






\begin{references}

\reference deBernardis, P. et al 2000, Nature 404, 955

\reference Brhlik, M., Chung, D. \& Kane, G. 2001, Intl. J. Mod. Phys D10, 367

\reference Burles, S., Nollett, K. \& Turner, M.S. 2001a, Phys. Rev. D 63, 063512

\reference Burles, S., Nollett, K. \& Turner, M.S. 2001b, ApJ 552, L1

\reference Carroll, S. 2001, http://www.livingreviews.org/Articles/Volume4/2001-1carroll

\reference Carroll, S. \& Kaplinghat, M. 2001, work in preparation

\reference Chaboyer, B. 2001, this volume

\reference Freedman, W. 2001, this volume

\reference Guth, A. 2001, this volume

\reference Hanany, S. et al 2000, ApJ 545, L5

\reference Hu, W., Sugiyama, N. \& Silk, J. 1997, Nature 386, 37

\reference Huterer, D. \& Turner, M.S. 2000, astro-ph/0012510 (Phys. Rev. D, in press)

\reference Jaffe, A. et al 2001, Phys. Rev. Lett. 86, 3475

\reference Jungman, G., Kamionkowski, M. \& Griest, K. 1996, Phys. Rep 267, 195

\reference Kolb, E.W. \& Turner, M.S. 1990, The Early Universe (Redwood City, CA:
Addison-Wesley)

\reference Kosowsky, A., Turner, M.S. \& Watkins, R. 1992,
Phys. Rev. Lett. 69, 2026

\reference Kragh, H. 1996, Cosmology and Controversy (Princeton, NJ:  Princeton Univ. Press)

\reference Krauss, L. \& Turner, M.S. 1999, Gen. Rel. Grav. 31, 1453

\reference Loeb, A. 1998, ApJ 499, L111

\reference Lopez, R. et al 1999, Phys. Rev. Lett. 82, 3952

\reference Netterfield, C.B. et al 2001, astro-ph/0104460

\reference O'Meara, J.M. et al 2001, ApJ 552, 718

\reference Olive, K., Steigman, G. \& Walker, T.P. 2000, Phys. Rep. 389,333

\reference Pryke, C. et al 2001, astro-ph/0104490 (submitted to ApJ)

\reference Schramm, D.N. \& Turner, M.S. 1998, Rev. Mod. Phys. 70, 303

\reference Tegmark, M. 2001, astro-ph/0101354 (submitted to Phys. Rev. D)

\reference Tonry, J. 2001, in this volume

\reference Turner, M.S. 1997a, in Generation of Cosmological Large-scale Structure,
eds. D.N. Schramm \& P. Galeotti (Dordrecht:  Kluwer), 153 (astro-ph/9704062)

\reference Turner, M.S. 1997b, Phys. Rev. D 55, R435

\reference Turner, M.S. 2000, Phys. Rep. 334, 619

\reference Turner, M.S. 2001a, PASP 113, 653

\reference Turner, M.S. 2001b, astro-ph/0106035 (submitted to ApJLett)

\end{references}
\end{document}